\documentclass[aps,prl,twocolumn,groupedaddress]{revtex4}

\usepackage{graphicx} 

\begin{document}

\renewcommand{\thefootnote}{\alph{footnote}}

\preprint{APL, Y. Takahashi}

\title{
Imaging of emission patterns in a T-shaped quantum wire laser
}

\author{
Yasushi Takahashi\footnote[1]{Electronic mail: taka8484@issp.u-tokyo.ac.jp}, 
Shinichi Watanabe,  
Masahiro Yoshita\footnote[2]{Also a visiting scientist at Bell laboratories, Lucent Technologies}, 
Hirotake Itoh, 
Yuhei Hayamizu, 
and Hidefumi Akiyama$^\dagger$}

\affiliation{
Institute for Solid State Physics, University of Tokyo, and CREST, JST,\\
5-1-5 Kashiwanoha, Kashiwa, Chiba 277-8581, Japan
}

\author{Loren N. Pfeiffer and Ken W. West}

\affiliation{Bell Laboratories, Lucent Technologies, 600 Mountain Avenue, Murray Hill, NJ 07974}

%\date{\today}
\date{March 14, 2003}

\begin{abstract}
Spatially and spectrally resolved microscopic images of spontaneous and stimulated emissions are imaged at the mirror facets of a GaAs T-shaped quantum wire laser with high uniformity. Laser emission from the one-dimensional ground state reveals a circular image located at the core of a T-shaped optical waveguide but significantly smaller in area than the low power spontaneous emission from the same waveguide. These images unambiguously allow assignment of all spontaneous and laser emissions to the wire ground state and respective intersecting wells in the structure.
\end{abstract}

\maketitle

Quantum wire lasers \cite{kapon1989,weg1993} should show outstanding performance because of enhanced density of states at the bottom of conduction and valence bands \cite{arakawa1982}.

A T-shaped quantum wire (T-wire) \cite{goni,someya} fabricated using cleaved-edge overgrowths \cite{loren1990} with molecular-beam epitaxy is advantageous, because it affords exact control of the one-dimensional (1D) structure. Stimulated emission from the 1D ground state was in fact first observed in a cleaved-edge overgrowth T-wire laser \cite{weg1993}. Photoluminescence (PL) images of the same T-wire laser were measured by Grober $et\ al$. with a low-temperature near-field scanning optical microscope (NSOM)\cite{grober1994}. They measured spatial distribution of three PL peaks and identified the PL peaks of T-wires and two adjacent wells. These measurements confirmed that one of the PL peaks evolving to the stimulated emission at higher pumping levels indeed comes from the T-wires region. However, the excitation level in that NSOM measurement was much lower than the lasing threshold, and the near-field emission patterns of lasing could not be measured. Such patterns, however, can be very important to the design of laser devices.

Another issue is that Rubio $et\ al$. observed a second lasing line L2 at high pumping levels in a similar T-wire structure \cite{rubio2001}. This L2 lasing occurred at higher energy than that of the 1D ground-state L1 lasing. The L2 lasing was suggested as possibly originating from an adjacent quantum well or from excited-states in T-wires. However, the actual origin of the L2 emission remains until now unanswered. For these reasons an experiment that could spatially image all of emissions in a T-wire laser would be important. 

We report, in this work, near-field emission patterns of a T-wire laser by microscopic imaging method with a 0.8 $\mu$m spatial resolution. We investigated a T-wire laser at 4 K from below threshold at low excitation power to a power 80 times higher than the threshold of L1 under optical pumping, where three lasing lines, L1, L2, and L3, are observed. 

\begin{figure}[b]
\centerline{\scalebox{0.4}{\includegraphics{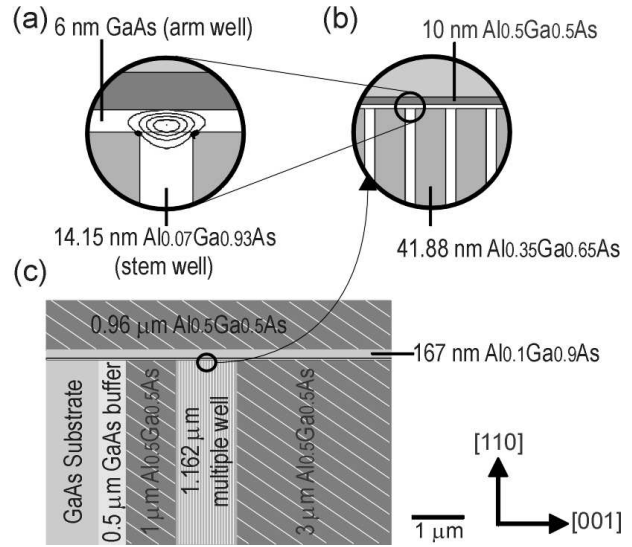}}}
\caption{Schematic cross section of T-wire laser structure. (a) T-wire at T-shaped intersection of stem well and arm well. Contour curves show constant probability ($|\psi|^2$ = 0.2, 0.4, 0.6, 0.8, 1.0) for electrons confined in a T-wire. (b) Multiple T-wire structure of stem wells and arm well. (c) All layer structure of T-wire laser sample.}
\end{figure}

Figure 1 shows a schematic view of our T-wire laser sample, where 20 T-wires are formed at T-shaped intersections of 20-period (001) multiple quantum wells (stem wells) and a (110) single quantum well (arm well). The 20 T-wires are embedded as shown in Fig. 1(c). It has a T-shaped optical waveguide core that is formed by 1.162 $\mu$m-thick multiple well layers with an averaged Al concentration of 28 \% and a 167 nm-thick $\rm{Al_{0.1}Ga_{0.9}As}$ layer, and this core is surrounded by $\rm{Al_{0.5}Ga_{0.5}As}$ optical cladding layers. The cavity length of the T-wire laser is 512 $\mu$m, and cavity facets are uncoated. The sample was fabricated in the same growth runs and has the same structure as that used in our previously reported samples \cite{akiyama2002}. Note that the interface uniformity of these T-wires is dramatically improved from earlier samples \cite{weg1993,goni,someya,grober1994,rubio2001} because we introduced growth-interrupt annealing at a high temperature in forming the arm well surface \cite{yoshita2002}. 

The T-wire laser was attached to a copper block and cooled down to 4 K. We used a continuous wave Titanium-Sapphire-laser to generate carriers in the active region. Excitation light was chopped into a 1\% duty ratio to minimize sample heating. The excitation was incident on the top (110) surface of the T-wire structure in a filament shape 2 $\mu$m in width and 700 $\mu$m in length through a 0.4 numerical aperture objective lens (N.A. : 0.4) and a 200-mm focal length cylindrical lens. The pumping energy was set at 1.648 eV so that the 167-nm $\rm{Al_{0.1}Ga_{0.9}As}$ layer was not excited, but was absorbed in the stem wells, the arm well, and the T-wires. Photo-excited carriers in the stem wells and the arm well diffuse into the lower energy T-wires. The configuration of excitation was optimized for minimizing the threshold of L1. The spontaneous and stimulated emissions that came out via one of the cavity-mirror surfaces was collected using a 0.5 numerical aperture objective lens (N.A. : 0.5). 

Figure 2 shows the emission spectra at three pumping levels 0.3, 6.5, and 210 mW incident on the (110) surface, measured with a spectral resolution of 0.2 meV. 

In spectrum (a) at an excitation level of 0.3 mW, three spontaneous emission peaks occur at 1.5804, 1.6030, and 1.6340 eV. As will be demonstrated below in the imaging experiment, these spontaneous emission peaks originate from the T-wires, arm well, and stem wells respectively.

In spectrum (b) at an excitation of 6.5 mW, a multi-mode lasing (L1) emission occurs at 1.5753 eV and becomes dominant over the other emissions. The threshold pumping power is about 5 mW, and the lasing energy is 5.1 meV below the spontaneous emission peak due to the T-wire ground-state exciton \cite{akiyama2002}. Note that the energy separation $\Delta E$ of the longitudinal modes is 0.27 meV. The effective refractive index $n_{\rm{eff}}$ estimated by $\Delta E = hc/2en_{\rm{eff}}L$ using a cavity length $L$ = 512 $\mu$m is 4.49. This value is very large compared with a refractive index 3.65 of GaAs. 

In spectrum (c) under a high excitation power of 210 mW, three lasing lines L1, L2, and L3 co-occur at 1.5736, 1.5883, and 1.6285 eV. Lasing threshold powers of L2 and L3 are about 150 and 80 mW. The spectrum shows that L1 lases in single-mode but that L2 and L3 are wide. The longitudinal modes of L3 are not seen, while those of L2 are resolved. The effective refractive index corresponding to the longitudinal mode separation of L2 is 4.51. As the excitation power further increases to 400 mW, the intensity of L2 overtakes that of L1 as reported with Rubio $et\ al$\cite{rubio2001}.

The energies of L1 and L3 are close to those of spontaneous emission peaks of the T-wires and the stem wells and suggest a strong link between them. However, the energy of L2 is not close to the spontaneous emission peak of the arm well. 

\begin{figure}
\centerline{\scalebox{0.4}{\includegraphics{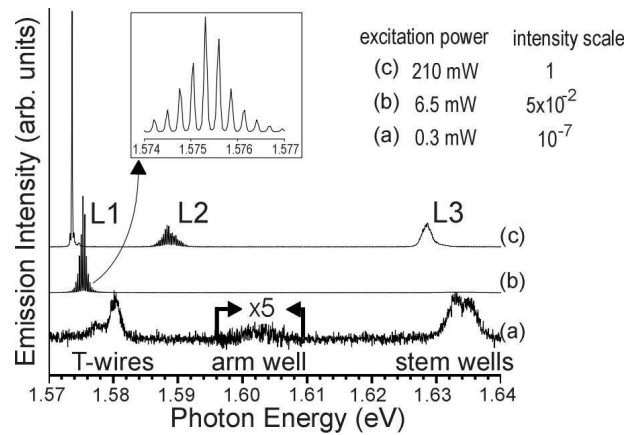}}}
\caption{Emission spectra of T-wire laser measured with spectral resolution of 0.2 meV at 4 K. (a) Three PL peaks from T-wires, arm well, and stem wells. (b) Lasing L1 from 1D ground state. (c)Coexistence of three lasing lines, L1$\sim$L3.}
\end{figure}

For these emission peaks, we measured spatially and spectrally resolved microscopic images using interference band-path filters. The images (b)$\sim$(g) in Fig. 3 show each emission pattern from a cavity mirror surface. The spatial and spectral resolutions are about 0.8 $\mu$m and 1 meV. White lines and cross-hatching superposed on each image show the $\rm{Al_{0.5}Ga_{0.5}As}$ cladding layers as shown in Fig. 1(c). The contour color in these images corresponds to the emission intensity as shown in the color bar. 

The black contour curves in Fig. 3(a) show the fundamental optical mode ($|\psi|^2$ = 0.9, 0.5, 0.1) at the T-shaped waveguide calculated using a finite element method. The superposed color image is a convolution of the optical mode with a 0.8 $\mu$m spatial resolution function. 

The images (b), (c), and (d) are spontaneous emission patterns at a low excitation level of 0.3 mW, which correspond to three PL peaks in Fig. 2(a) to be assigned to T-wires, arm well, and stem wells, respectively.

The image (b) of the low-energy peak at 1.5804 eV is consistent with the assignment of its origin to the T-wires because the center of the emission exists at the T-shaped waveguide containing 20 T-wires. Though the emission pattern is slightly flat near (110) surface, it is very similar to the convoluted image (a) of the T-shaped waveguide mode. 

\begin{figure}
\centerline{\scalebox{0.4}{\includegraphics{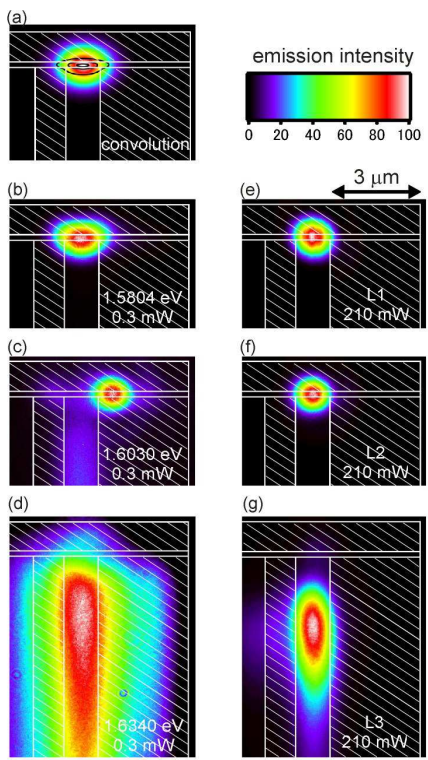}}}
\caption{Images of emission patterns in T-wire laser at 4 K. Cross-hatching indicate $\rm{Al_{0.5}Ga_{0.5}As}$ cladding. (a) Convolution image of calculated optical mode at the T-shaped waveguide. (b)-(g) Microscopic images of emission patterns measured with spatial resolution of 0.8 $\mu$m.}
\end{figure}

The emission pattern (c) of the middle-energy peak at 1.6030 eV shows that the center of the emission is located at the 3 $\mu$m-wide region of the arm well on the $\rm{Al_{0.5}Ga_{0.5}As}$ cladding layer (see Fig. 1(c)). The 1 $\mu$m-wide region near the substrate shows weak emission intensity because the carriers tend to flow away into T-wires and a GaAs buffer layer. The arm well between the T-wires in the core region on the $\rm{Al_{0.35}Ga_{0.65}As}$ barrier also shows very weak emission intensity because carriers in this region mostly flow into adjacent T-wires. 

The emission pattern (d) of the high-energy peak at 1.6340 eV shows that the emission originates from stem wells. Decreased intensity at the region away from the T-wires is most likely due to decreased excitation light by absorption and indicates the penetration depth. Note that the emission of the stem wells is quenched in the region near the T-wires due to diffusion of carriers into the T-wires \cite{grober1994}. The majority of the carriers in the T-wires are supplied in this mechanism because the volume of the stem wells in this region is larger than that of the T-wires and the arm well. These carrier diffusion from the arm well and the stem wells into the T-wires and the volume of each structure roughly explain the relative intensities of the three PL peaks in spectrum (a) in Fig. 2.

The images (e), (f), and (g) are laser emission patterns of L1, L2, and L3 at a high excitation level of 210 mW. 

The emission pattern (e) of L1 is centered at the peak of PL patterns (b) of the T-wires. L1 is 5$\sim$7 meV below the exciton peak of the T-wire ground state, and it justifies the assignment that L1 is the lasing from the ground state of the T-wires. Similarly, emission pattern (g) of L3 demonstrates the assignment that L3 is the lasing from the stem wells. On the other hand, emission pattern (f) of L2 is not centered at the peak of PL patterns (c) of the arm well but has almost the same pattern of L1. This proves that the arm well in the cladding region is not the origin of L2. Instead, L2 should be ascribed to the arm well between the T-wires in the core region or some excited states of the T-wire. 

To further discuss the origin of L2, we now refer to an additional experiment, where we made a similar lasing measurement on a single T-wire laser sample with an uncoated-facet cavity of 500 $\mu$m length at 4 K. The single wire laser sample has the same structure as that reported by Hayamizu $et\ al$\cite{hayamizu2002} except for the uncoated facet mirrors. It has only one T-wire defined by a 14 nm stem well and a 6 nm arm well. If we compare the single wire laser with the 20-wire laser, the total volume of the T-wire is reduced by a factor of 20, while that of the arm well is almost unchanged. Thus, the gain due to the T-wire ground state and the T-wire excited states should be reduced by a factor of 20, while the gain due to the arm well should be almost unchanged. The results of the experiment on the single wire laser show that L1 is not observed, while L2 is observed with a similar lasing-threshold to that of the 20-wire laser. This experimental fact suggests that the origin of L2 is most probably the arm well with a similar total volume.

It is interesting, in particular, that the emission patterns of L1 and L2 are small and have nearly a circular shape. They are smaller than the convolution image (a) of the T-shaped waveguide mode defined by the background refractive index. The emission pattern of L1 remains unchanged and stable against variation on excitation powers from just above the threshold to 400 mW. Even if the stripe width of the pumping laser is broadened to excite 20 T-wires more uniformly or if the position of the excitation light is moved from the center of 20 T-wires, the pattern of L1 is the same. A circular near-field emission pattern indicates an important advantage in laser applications, because it results in a symmetric far-field beam pattern without adding correction optics. 

The mechanism of this small and circular near-field emission pattern is not understood. This may have some relation with the high effective refractive index of 4.5 obtained from the longitudinal-mode separation of L1 and L2 found in Fig. 2. Carriers causing optical gain in T-wires might enhance the effective refractive index at the center of the core region and modify both longitudinal and horizontal modes of lasing. 

In summary, three lasing lines are observed under optical pumping in a T-wire laser with high uniformity at 4 K. We measured microscopic images of each emission peak. Origins of the three lasing lines are assigned: low-energy lasing L1 is due to the T-wire ground state, middle-energy lasing L2 is most probably due to the core-arm well, and high-energy lasing L3 is due to the stem wells. The near-field emission pattern of L1 is small and has a circular shape, which does not change up to an excitation power 80 times higher than the lasing threshold of L1. The emission pattern of L2 is almost the same as that of L1. A large effective refractive index of 4.5 is found for L1 and L2, which may be related to the circular emission patterns. 

This work is partly supported by a Grant-in-Aid from the MEXT, Japan.

\end{document}